\documentstyle[12pt]{article}

\begin{document}
\thispagestyle{empty}

\begin{center}
\LARGE \tt \bf{Stretch fast dynamo mechanism via conformal mapping
in Riemannian manifolds}
\end{center}

\vspace{1.5cm}

\begin{center} {\large L.C. Garcia de Andrade\footnote{Departamento de
F\'{\i}sica Te\'{o}rica - Instituto de F\'{\i}sica - UERJ

Rua S\~{a}o Fco. Xavier 524, Rio de Janeiro, RJ

Maracan\~{a}, CEP:20550-003 , Brasil.E-mail:garcia@dft.if.uerj.br.}}
\end{center}

\vspace{2.0cm}

\begin{abstract}
Two new analytical solutions of self-induction equation, in
Riemannian manifolds are presented. The first represents a twisted
magnetic flux tube or flux rope in plasma astrophysics, which shows
that the depending on rotation of the flow the poloidal field is
amplified from toroidal field which represents a dynamo. The value
of the amplification depends on the Frenet torsion of the magnetic
axis of the tube. Actually this result illustrates the Zeldovich
stretch, twist and fold (STF) method to generate dynamos from
straight and untwisted ropes. Motivated by the fact that this
problem was treated using a Riemannian geometry of twisted magnetic
flux ropes recently developed (Phys Plasmas (2006)), we investigated
a second dynamo solution which is conformally related to the Arnold
kinematic fast dynamo. In this solution it is shown that the
conformal effect on the fast dynamo metric only enhances the
Zeldovich stretch, and therefore a new dynamo solution is obtained.
When a conformal mapping is performed in Arnold fast dynamo line
element a uniform stretch is obtained in the original line element.
 \vspace{0.5cm}
\noindent {\bf PACS
numbers:\hfill\parbox[t]{13.5cm}{02.40.Hw-Riemannian geometries}}
\end{abstract}
\newpage
\section{Introduction}
 Geometrical tools have been used with success \cite{1} in
 Einstein general relativity (GR) have been also used
 in other important areas of physics, such as plasma structures in tokamaks
 as been clear in the Mikhailovskii \cite{2} book to investigate the tearing
 and other sort of instabilities in confined plasmas \cite{2}, where the Riemann
 metric tensor plays a dynamical role interacting with the magnetic field
 through the magnetohydrodynamical equations (MHD). Recently
 Garcia de Andrade \cite{3,4} has also made use of Riemann metric to investigate
 magnetic flux tubes in superconducting plasmas. Thiffault and Boozer \cite{5} following the same reasoning applied the methods of Riemann
 geometry in the context of chaotic flows and fast dynamos. Yet more recently Thiffeault \cite{6} investigated the stretching and
 Riemannian curvature of material lines in chaotic flows as possible dynamos models. An interesting tutorial review of chaotic flows
 and kinematical dynamos has been presented earlier by Ott \cite{7}. Also Boozer \cite{8} has obtained a geomagnetic dynamo from conservation
 of magnetic helicity. Actually as pointed out by Baily and Childress \cite{9} have called the attention to the fact that
 the focus nowadays the focus of kinematic dynamo theory is on the fast dynamos, specially in construction on curved
 Riemannian manifolds to stretch, twist and fold magnetic flows filaments or tubes to generated dynamo solutions. This
 method was invented by Zeldovich \cite{10}. In this paper taking the advantage of the success of conformal Riemannian
 geometry techniques used to find out new solutions of Einstein general relativistic field equations\cite{1},
 we use this same conformal geometrical technique to find new solutions of the
incompressible flows in Arnold metric \cite{11,12,13}.
 To resume , finding and being able to recognize the existence of
 dynamos or non-dynamos is not only important from physical and mathematical
 point of view, but also finding mathematical techniques which
 allows us to obtain new dynamo solution from dynamos or either
 nondynamo metrics such as the one we consider here as a flavour of
 the difficult to find a dynamo solution in Riemannian manifolds.
 This also motivates us to show that conformally Arnold fast dynamo
 metric, is also a dynamo solution. Recently Hanasz and Lesch \cite{14} have used also a conformal Riemannian metric in
 ${\cal{E}}^{3}$
 to investigate the galactic dynamo also using magnetic flux tubes. Also recently, Kambe \cite{15} and
 Hattori and Zeitlin \cite{16}, have investigated the rate of stretching of Riemannian line elements
 of imcompressible fluids, in the framework of differential geometry
 of diffeomorphisms, in the spirit of Zeldovich stretching. Actually
they considered the exponential
 stretching of line elements in time, or dynamo action, in the context of negative curvature in turbulent flows. They also
 considered the concentration of vortex and magnetic flux tube. This provides a strong physical motivation to the
 present investigation. The paper is organized as follows: In section II the
 the dynamo Riemann metric representing flux rope analytical solution of the self-induction equation in the
 case or zero resistivity is presented. In section III the dynamo solution in Riemannian conformal geometry
 is given. In section IV Riemannian curvature of a particular conformal dynamo is computed and in section V conclusions
 are presented.
 \section{Thin flux rope dynamos in Riemannian manifold}
In this section we shall consider generalization of the Riemann
metric of a stationary twisted magnetic flux tube as considered by
Ricca \cite{17} in the Riemann manifold to address the nonstationary
case where the toroidal and poloidal magnetic fields, in principle,
may depend on time. With this metric at hand , we are to solve
analytical the self-induction magnetic flow equation to check for
the dynamo existence.  Let us now start by considering the MHD field
equations
\begin{equation}
{\nabla}.\vec{B}=0 \label{1}
\end{equation}
\begin{equation}
\frac{{\partial}}{{\partial}t}\vec{B}-{\nabla}{\times}[\vec{u}{\times}\vec{B}]-{\eta}{\nabla}^{2}\vec{B}=0
 \label{2}
\end{equation}
where $\vec{u}$ is a solenoidal field while ${\eta}$ is the
diffusion coefficient. Equation (\ref{2}) represents the
self-induction equation. The vectors $\vec{t}$ and $\vec{n}$ along
with binormal vector $\vec{b}$ together form the Frenet frame which
obeys the Frenet-Serret equations
\begin{equation}
\vec{t}'=\kappa\vec{n} \label{3}
\end{equation}
\begin{equation}
\vec{n}'=-\kappa\vec{t}+ {\tau}\vec{b} \label{4}
\end{equation}
\begin{equation}
\vec{b}'=-{\tau}\vec{n} \label{5}
\end{equation}
the dash represents the ordinary derivation with respect to
coordinate s, and $\kappa(s,t)$ is the curvature of the curve where
$\kappa=R^{-1}$. Here ${\tau}$ represents the Frenet torsion. The
gradient operator becomes
\begin{equation}
{\nabla}=\vec{t}\frac{\partial}{{\partial}s}+\vec{e_{\theta}}\frac{1}{r}\frac{\partial}{{\partial}{\theta}}+
\vec{e_{r}}\frac{\partial}{{\partial}r} \label{6}
\end{equation}
 Now we shall consider the analytical solution of the self-induction magnetic equation investigated which represents a
 non-dynamo thin magnetic flux rope. Before the derivation of this result , we would like to point it out that it is not
 trivial, since the Zeldovich antidynamo theorem states that the two dimensional magnetic fields do not support
 dynamo action, and here as we shall see bellow, the flux tube axis possesses Frenet curvature as well as torsion and this
 last one cannot take place in planar curves. Let us now consider here the metric of magnetic
 flux tube
\begin{equation}
ds^{2}=dr^{2}+r^{2}d{{\theta}_{R}}^{2}+{K^{2}}(s)ds^{2} \label{7}
\end{equation}
This is a Riemannian line element
\begin{equation}
ds^{2}=g_{ij}dx^{i}dx^{j} \label{8}
\end{equation}
if the tube coordinates are $(r,{\theta}_{R},s)$ \cite{15} where
${\theta}(s)={\theta}_{R}-\int{{\tau}ds}$ where $\tau$ is the Frenet
torsion of the tube axis and $K(s)$ is given by
\begin{equation}
{K^{2}}(s)=[1-r{\kappa}(s)cos{\theta}(s)]^{2} \label{9}
\end{equation}
Since we are considered thin magnetic flux tubes, this expression is
$K\approx{1}$ in future computations. Computing the Riemannian
Laplacian operator ${\nabla}^{2}$ in curvilinear coordinates
\cite{16} one obtains
\begin{equation}
{\nabla}^{2}=\frac{1}{\sqrt{g}}{\partial}_{i}[\sqrt{g}g^{ij}{\partial}_{j}]
\label{10}
\end{equation}
where ${\partial}_{j}:=\frac{{\partial}}{{\partial}x^{j}}$ and
$g:=det{g_{ij}}$ where $g_{ij}$ is the covariant component of the
Riemann metric of the flux rope. Here, to better compare the dynamo
action generation of toroidal field from poloidal fields we shall
consider the that the toroidal component of magnetic field
$B_{s}(s)$ is given in the Frenet frame as
\begin{equation}
\vec{B_{s}}=b_{0}(s)\vec{t} \label{11}
\end{equation}
Note also that we have considered that the flux rope magnetic field
does not depend on the r and ${\theta}_{R}$ coordinates. While the
poloidal magnetic field the magnetic field here can be expressed as
\begin{equation}
\vec{B_{\theta}}(t,{\theta})=e^{pt}b_{1}\vec{e_{\theta}} \label{12}
\end{equation}
Now let us substitute the definition of the poloidal plus toroidal
magnetic fields into the self-induction equation, which with the
help of the expressions
\begin{equation}
\vec{e_{\theta}}=-\vec{n}sin{\theta}+\vec{b}cos{\theta} \label{13}
\end{equation}
\begin{equation}
{{\partial}_{\theta}}\vec{e_{\theta}}=-\vec{n}[(1+{\tau}^{-1}\kappa){sin{\theta}}+cos{\theta}]-\vec{b}[cos{\theta}+sin{\theta}]
\label{14}
\end{equation}
and
\begin{equation}
{\partial}_{t}\vec{e_{\theta}}={\omega}\vec{e}_{\theta}-{\partial}_{t}\vec{n}sin{\theta}+{\partial}_{t}\vec{b}cos{\theta}\label{15}
\end{equation}
Considering the equations for the time derivative of the Frenet
frame given by the hydrodynamical absolute derivative
\begin{equation}
\dot{\vec{X}}={\partial}_{t}\vec{X}+[\vec{v}.{\nabla}]\vec{X}
\label{16}
\end{equation}
where $\vec{X}=(\vec{t},\vec{n},\vec{b})$ represents the Frenet
frame into the expressions for the total derivative of each  Frenet
frame vectors
\begin{equation}
\dot{\vec{t}}={\partial}_{t}\vec{t}+[{\kappa}'\vec{b}-{\kappa}{\tau}\vec{n}]
\label{17}
\end{equation}
\begin{equation}
\dot{\vec{n}}={\kappa}\tau\vec{t} \label{18}
\end{equation}
\begin{equation}
\dot{\vec{b}}=-{\kappa}' \vec{t} \label{19}
\end{equation}
one obtains the values of respective partial derivatives of the
Frenet frame as
\begin{equation}
{\partial}_{t}\vec{t}=-{\tau}{\kappa}[1-{\kappa}{\tau}^{-2}\frac{v_{\theta}}{r}]\vec{n}
\label{20}
\end{equation}
\begin{equation}
{\partial}_{t}{\vec{n}}={\tau}{\kappa}[1-{\kappa}\vec{\tau}^{-2}\frac{v_{\theta}}{r}]\vec{t}+\frac{v_{\theta}}{r}\vec{b}
\label{21}
\end{equation}
\begin{equation}
{\partial}{\vec{b}}={\kappa}{\tau}^{-1}\frac{v_{\theta}}{r}\vec{n}
\label{22}
\end{equation}
where we have used the hypothesis that $\dot{\vec{b}}=0$ or
${\kappa}'(t,s)=0$, which means that the curve curvature only
depends on time. A simple example from solar physics, would be a
flux tube curved and with torsion oscillating with fixed sun spots.
Substitution of these vectorial expressions into expression
(\ref{15}) yields
\begin{equation}
{\partial}_{t}\vec{e_{\theta}}=[{\omega}cos{\theta}-\frac{v_{\theta}}{r}]\vec{b}-[{\omega}sin{\theta}-{\tau}^{-1}{\kappa}\frac{v_{\theta}}{r}]\vec{n}
+ [\kappa\tau(1+{\tau}^{-2}\frac{v_{\theta}}{r})sin{\theta}]\vec{t}
\label{23}
\end{equation}
along with the equation
\begin{equation}
\frac{{\partial}B_{\theta}}{{\partial}s}=B_{\theta}r\tau\kappa
\label{24}
\end{equation}
and the fact that $\frac{{\partial}B_{s}}{{\partial}s}=0$, together
with the self-induction equation we obtain the following system of
equations, for a highly conductive fluid as our own universe, with
resistivity ${\eta}=0$
\begin{equation}
{\partial}_{t}B_{\theta}+{\tau}v_{\theta}sin{\theta}B_{\theta}=0
\label{25}
\end{equation}
\begin{equation}
sin{\theta}B_{\theta}+\frac{B_{s}}{sin{\theta}}[1+{\tau}^{-2}\frac{v_{\theta}}{r}]=0
\label{26}
\end{equation}
To obtain these last two expressions we assume that
$v_{\theta}>>>v_{s}$ and that ${\partial}_{s}v{s}=0$ and that the
continuity equation
\begin{equation}
{\nabla}.\vec{v}=0 \label{27}
\end{equation}
where we have considered that the flow is imcompressible which is a
reasonable approximation in plasma physics. This expresion yields
\begin{equation}
\frac{{\partial}v_{\theta}}{{\partial}s}+v_{\theta}r\tau\kappa=0
\label{28}
\end{equation}
Equation (\ref{25}) can be rewritten as
\begin{equation}
[p+{\tau}v_{\theta}sin{\theta}]=0 \label{29}
\end{equation}
which upon substitution on the equation (\ref{26}) yields
\begin{equation}
\frac{B_{\theta}}{B_{s}}{sin{\theta}}[1+{\tau}^{-2}\frac{v_{\theta}}{r}]=0
\label{30}
\end{equation}
which with the assumption that the flux tube has a small twist and
${v_{\theta}}^{2}<<1$. Now these equations
\begin{equation}
\frac{B_{\theta}}{B_{s}}=\frac{{\tau}{\omega}r}{p^{2}} \label{31}
\end{equation}
since $B_{s}\approx{constant}b_{0}$ by hypothesis, we have that
relation (\ref{31}) tells us that the relation
$\frac{{\tau}{\omega}r}{p^{2}}>0$ implies that the poloidal field is
amplified from the constant modulus toroidal field, which is the
dynamo rope condition. This physical situation happens often in the
sun. Thus relation between the angular flow speed and the torsion
and radius distance shows that there is a lower bound for the thin
rope dynamo which is given by $r>\frac{p^{2}}{{\omega}{\tau}}$. The
divergence-free equations for the magnetic and flow fields, allows
us to write down the solutions for $B_{\theta}$
\begin{equation}
{B_{\theta}}=B^{0}e^{pt-\int{\frac{r}{R}cos{\theta}d{\theta}}}
\label{32}
\end{equation}
which if we recall the definition of the deviation of flat Riemann
metric $K$ of the tube above , we may express the integral in terms
of $K(s)$ as
\begin{equation}
{B_{\theta}}=B^{0}e^{pt-\int{(1-K(s))d{\theta}}} \label{33}
\end{equation}
which shows that in the very thin flux rope dynamo in this solution
the effect curvature of the Riemannian tube is minor. Here we have
used the fact that the external curvature of the rope is given by
${\kappa}_{0}=\frac{1}{\kappa}_{0}$. The solution for torsion and
velocity flow are essentially analogous.
\section{Conformal dynamos on manifolds}
Conformal mapping on a Riemannian line element can in general be
represented by
\begin{equation}
ds^{2}=e^{2{\lambda}_{0}(\vec{x})}(ds_{0})^{2} \label{34}
\end{equation}
Manifolds related in this manner, are said conformally related. Note
also that this is intrinsically connected to the stretch part of STF
mechanism to generate rope dynamos. Actually in the previous section
we considered a Riemannian metric for twisted flux tube which was
stretched solely on the ds element along the magnetic axis of the
dynamo rope, through the factor $K(s)$, of course when the flux rope
is thin, the this stretch effect almost vanish though twist and fold
may still be kept. Therefore strictly speaking, this is not a
conformal mapping but only a stretch, therefore strictly speaking
not all stretches are represented by conformal mapping but every
conformal metric represents stretching in the Riemannian manifold.
One of disadvantages of conformal stretching is that the stretching
is uniform as in the case of Arnold fast dynamo metric. Conformal
metric techniques have also been widely used as a powerful tool
obtain new solutions of the Einstein's field equations of GR from
known solutions. By analogy, here we are using this method to yield
new solutions of MHD dynamo from the well-known fast dynamo Arnold
solution. We shall demonstrate that distinct physical features from
the Arnold solution maybe obtained. Before that we just very briefly
review the Arnold solution. The Arnold metric line element can be
defined as \cite{11}
\begin{equation}
ds^{2}=e^{-2{\lambda}z}dp^{2}+e^{2{\lambda}z}dq^{2}+dz^{2}
\label{35}
\end{equation}
which describes a dissipative dynamo model on a 3D Riemannian
manifold. By dissipative here, we mean that contrary to the previous
section, the resistivity $\eta$ is small but finite. The flow build
on a toric space in Cartesian coordinates $(p,q,z)$ given by
$T^{2}\times[0,1]$ of the two dimensional torus. The coordinates p
and q are build as the eigenvector directions of the toric cat map
in ${\cal R}^{3}$ which possesses eigenvalues as
${\chi}_{1}=\frac{(3+\sqrt{5})}{2}>1$ and
${\chi}_{2}=\frac{(3-\sqrt{5})}{2}<1$ respectively. Note for example
that if perform a simple constant conformal mapping such that
\begin{equation}
{ds}^{2}=e^{2({\lambda}z+{\lambda}_{0})}{dp}^{2}+e^{-2({\lambda}z-{\lambda}_{0})}{dq}^{2}+e^{2{\lambda}_{0}}dz^{2}\label{36}
\end{equation}
which represents a simple global translation and is not changed at
every point in the manifold. Let us now recall the Arnold et al
\cite{13} definition of a orthogonal basis in the Riemannian
manifold ${\cal{M}}^{3}$
\begin{equation}
\vec{e}_{p}=e^{{\lambda}z}\frac{{\partial}}{{\partial}p}\label{37}
\end{equation}
\begin{equation}
\vec{e}_{q}=e^{-{\lambda}z}\frac{{\partial}}{{\partial}q}\label{38}
\end{equation}
\begin{equation}
\vec{e}_{z}=\frac{{\partial}}{{\partial}q}\label{39}
\end{equation}
Assume a magnetic vector field $\vec{B}$ on M
\begin{equation}
\vec{B}=B_{p}\vec{e}_{p}+B_{q}\vec{e}_{q}+B_{z}\vec{e}_{z}\label{40}
\end{equation}
The vector analysis formulas in this frame are
\begin{equation}
{\nabla}f=[e^{{\lambda}z}{\partial}_{p}f,e^{-{\lambda}z}{\partial}_{q}f,{\partial}_{z}f]\label{41}
\end{equation}
where f is the map function $f:{\cal{R}}^{3}\rightarrow{\cal{R}}$.
The Laplacian is given by
\begin{equation}
{\Delta}f={\nabla}^{2}f=[e^{2{\lambda}z}{{\partial}_{p}}^{2}f+e^{-2{\lambda}z}{{\partial}_{q}}^{2}f+{{\partial}_{z}}^{2}f]
\label{42}
\end{equation}
while the divergence is given by
\begin{equation}
{\nabla}.\vec{B}=div\vec{B}=div[B_{p}\vec{e}_{p}+B_{q}\vec{e}_{q}+B_{z}\vec{e}_{z}]=[e^{{\lambda}z}{{\partial}_{p}}B_{p}+
e^{-{\lambda}z}{{\partial}_{q}}B_{q}+{{\partial}_{z}}B_{z}]\label{43}
\end{equation}
In particular one may write
\begin{equation}
div{\vec{e}}_{p}=div{\vec{e}}_{q}=div{\vec{e}}_{z}=0\label{44}
\end{equation}
in turn the curl is written as
\begin{equation}
curl\vec{B}=curl[B_{p}\vec{e}_{p}+B_{q}\vec{e}_{q}+B_{z}\vec{e}_{z}]\label{45}
\end{equation}
where
\begin{equation}
curl_{p}\vec{B}=e^{-{\lambda}z}({\partial}_{q}B_{z}-{\partial}_{z}(e^{{\lambda}z}B_{q}))\label{46}
\end{equation}
\begin{equation}
curl_{q}\vec{B}=-e^{{\lambda}z}({\partial}_{p}B_{z}-{\partial}_{z}(e^{-{\lambda}z}B_{p}))\label{47}
\end{equation}
\begin{equation}
curl_{z}\vec{B}=e^{{\lambda}z}{\partial}_{p}B_{q}-e^{-{\lambda}z}{\partial}_{q}B_{p}\label{48}
\end{equation}
and
\begin{equation}
curl{\vec{e}}_{p}=-{\lambda}{\vec{e}}_{q}\label{49}
\end{equation}
\begin{equation}
curl{\vec{e}}_{q}=-{\lambda}{\vec{e}}_{p}\label{50}
\end{equation}
\begin{equation}
curl{\vec{e}}_{z}=0\label{51}
\end{equation}
The Laplacian operators of the frame basis are
\begin{equation}
{\Delta}{\vec{e}}_{p}=-curlcurl{\vec{e}}_{p}=-{\lambda}^{2}{\vec{e}}_{p}
\label{52}
\end{equation}
\begin{equation}
{\Delta}{\vec{e}}_{q}=-curlcurl{\vec{e}}_{q}=-{\lambda}^{2}{\vec{e}}_{q}
\label{53}
\end{equation}
\begin{equation}
{\Delta}{\vec{e}}_{z}=0 \label{54}
\end{equation}
from these expressions Arnold et al \cite{13} were able to build the
self-induced equation in this Riemannian manifold as
\begin{equation}
{\partial}_{t}B_{p}+v{\partial}_{z}B_{p}=-{\lambda}vB_{p}+{\eta}[{\Delta}-{\lambda}^{2}]{B}_{p}-
2{\lambda}e^{{\lambda}z}{\partial}_{p}B_{z} \label{55}
\end{equation}
\begin{equation}
{\partial}_{t}B_{q}+v{\partial}_{z}B_{q}=+{\lambda}vB_{q}+{\eta}[{\Delta}-{\lambda}^{2}]{B}_{p}-
2{\lambda}e^{-{\lambda}z}{\partial}_{q}B_{z} \label{56}
\end{equation}
\begin{equation}
{\partial}_{t}B_{z}+v{\partial}_{z}B_{z}={\eta}[{\Delta}-2{\lambda}{\partial}_{z}]B_{z}
\label{57}
\end{equation}
Decomposing the magnetic field on a Fourier series, Arnold et al
were able to yield the following solution
\begin{equation}
b(p,q,z.t)=e^{{\lambda}vt}b(p,q,z-vt,0) \label{58}
\end{equation}
where $B(x,y,z,t)=b(p,q,z,t)$ and the fast dynamo limit $\eta=0$ was
used. Now with these formulas , we are able to compute the solution
of the self-induced magnetic equation in the background of conformal
Riemannian line element
\begin{equation}
{ds}^{2}={\Omega}(z)[e^{-2{\lambda}z}{dp}^{2}+e^{2{\lambda}z}{dq}^{2}+dz^{2}]\label{59}
\end{equation}
The reason for using the general conformal stretching factor
${\Omega}(z)$ instead of the previous exponential stretching , is to
show that the dynamo obtained is not only due to the exponential
stretching but conformal dynamos, allow for the existence of more
general conformal stretching. A far obvious, though important
observation here is the fact that from the equations bellow, we
recover the Arnold et al \cite{13} if we simply make the conformal
factor ${\Omega}:=1$. Denoting the dual one form for Arnold basis as
\begin{equation}
{\phi}_{p}=e^{-{\lambda}z}dp \label{60}
\end{equation}
\begin{equation}
{\phi}_{q}=e^{{\lambda}z}dq \label{61}
\end{equation}
\begin{equation}
{\phi}_{z}=dz \label{62}
\end{equation}
one obtains the Arnold fast dynamo metric as
\begin{equation}
ds^{2}={{\phi}_{p}}^{2}+{{\phi}_{q}}^{2}+{{\phi}_{z}}^{2} \label{63}
\end{equation}
Thus the conformal one form dual basis can be expressed as
\begin{equation}
{{\phi}_{p}}^{C}={\Omega}^{\frac{1}{2}}{\phi}_{p} \label{64}
\end{equation}
\begin{equation}
{{\phi}_{q}}^{C}={\Omega}^{\frac{1}{2}}{\phi}_{q} \label{65}
\end{equation}
\begin{equation}
{{\phi}_{z}}^{C}={\Omega}^{\frac{1}{2}}{\phi}_{z} \label{66}
\end{equation}
On the other hand the vector field basis in conformal metric becomes
\begin{equation}
\vec{e}_{p}={\Omega}^{-\frac{1}{2}}e^{{\lambda}z}\frac{{\partial}}{{\partial}p}\label{67}
\end{equation}
\begin{equation}
\vec{e}_{q}={\Omega}^{-\frac{1}{2}}e^{-{\lambda}z}\frac{{\partial}}{{\partial}q}\label{68}
\end{equation}
\begin{equation}
\vec{e}_{z}={\Omega}^{-\frac{1}{2}}\frac{{\partial}}{{\partial}z}\label{69}
\end{equation}
Let us now repeat some of the fundamental vector analysis relations
above in the conformal geometry. The first is the Laplacian of
$\vec{B}$
\begin{equation}
{\Delta}_{C}\vec{B}={\Omega}^{-1}{\Delta}\vec{B}-\frac{1}{2}{\Omega}^{-2}[{\partial}_{z}{\Omega}]{\partial}_{z}\vec{B}\label{70}
\end{equation}
A fundamental change here in the conformal stretching in Riemannian
geometrical dynamos, is that of the velocity flow. In Arnold fast
dynamo example, the flow is a very simple one which is given by
$(0,0,v)$ where v is constant. Here the dynamo flow is effectively
in the dynamo equation by a term of the form
\begin{equation}
({\vec{v}}.{\nabla}){B}_{p}={\Omega}^{-1}v{\partial}_{z}B_{p}\label{71}
\end{equation}
which shows that a nonconstant effective velocity flow such as
$v_{eff}={\Omega}^{-1}v$ would act in conformal dynamos with respect
to the previous Arnold example. The other fundamental component of
the $curl[\vec{v}{\times}\vec{B}]$ given by
\begin{equation}
({\vec{B}}.{\nabla})\vec{v}={\Omega}^{-2}v[{\partial}_{z}{\Omega
}]{\vec{e}}_{z}\label{72}
\end{equation}
With these expressions from conformal geometry in hand, we are now
able to express the Arnold et al dynamo equations are
\begin{equation}
{\partial}_{t}\vec{B}+(\vec{v}.{\nabla}_{C})\vec{B}=(\vec{B}.{\nabla}_{C})\vec{v}+{\eta}{\Delta}_{C}\vec{B}
\label{73}
\end{equation}
In terms of components this conformal self-induced equation in the
Riemannian manifold can be expressed as
\begin{equation}
{\partial}_{t}B_{p}+{\Omega}^{-1}v{\partial}_{z}B_{p}=-{\lambda}{\Omega}^{-1}vB_{p}+{\eta}[({\Delta}-{\lambda}^{2})B_{p}-
2{\lambda}e^{{\lambda}z}{\partial}_{p}B_{z}]-\frac{\eta}{2}{\Omega}^{-2}({\partial}_{z}{\Omega})({\partial}_{z}B_{p}+{\lambda}B_{p})
\label{74}
\end{equation}
The equation for q component can be obtained from p one by simply
performing the substitution ${\lambda}\rightarrow{-{\lambda}}$. The
expression for component-z is
\begin{equation}
{\partial}_{t}B_{z}+{\Omega}^{-1}v{\partial}_{z}B_{z}={\eta}[{\Delta}-2{\lambda}{\partial}_{z}-\frac{1}{2}{\Omega}^{-2}{\partial}_{z}{\Omega}{\partial}_{z}]B_{z}
\label{75}
\end{equation}
Decomposing again the magnetic field on a Fourier series now in
conformal geometry, yields the following solution
\begin{equation}
b(p,q,z.t)=e^{{\lambda}{\Omega}^{-1}vt}b(p,q,z-vt,0)=e^{{\lambda}v_{eff}t}b(p,q,z-vt,0)
\label{76}
\end{equation}
where is the conformal Riemannian fast dynamo solution. As given
explicitly in this solution the basic effect of the conformal
geometry in fast dynamos is on the speed of dynamo which is an
important physical effect.
\section{Riemann curvature of conformal dynamos}
The important role of negative curvature of geodesic flows in
dynamos have been investigated by Anosov \cite{19}. These Anosov
flows, though somewhat artificial, provide excelent examples for
numerical computation experiments in fast dynamos \cite{20}. Within
this motivation we include here a simple example of conformal
spatially stretching, where ${\Omega}:= e^{1{\lambda}z}$. This
conformal stretching applied to Arnold metric yields the conformal
metric as
\begin{equation}
ds^{2}={dp}^{2}+e^{4{\lambda}z}{dq}^{2}+e^{{\lambda}z}dz^{2}\label{77}
\end{equation}
or in terms of the frame basis form ${\omega}^{i}$ $(i=1,2,3)$ is
\begin{equation}
ds^{2}=({{\omega}^{p}})^{2}+({{\omega}^{q}})^{2}+({{\omega}_{z}})^{2}\label{78}
\end{equation}
The basis form are write as
\begin{equation}
{\omega}^{p}=dp \label{79}
\end{equation}
\begin{equation}
{\omega}^{q}=e^{{\lambda}z}dq \label{79}
\end{equation}
and
\begin{equation}
{\omega}^{z}=e^{{\frac{\lambda}{2}}z}dq \label{80}
\end{equation}
By applying the exterior differentiation in this basis form one
obtains
\begin{equation}
d{\omega}^{p}=0 \label{81}
\end{equation}
\begin{equation}
d{\omega}^{z}=0 \label{82}
\end{equation}
and
\begin{equation}
d{\omega}^{q}={\lambda}e^{-{\frac{\lambda}{2}}z}{\omega}^{z}{\wedge}{\omega}^{q}
\label{83}
\end{equation}
Substitution of these expressions into the first Cartan structure
equations one obtains
\begin{equation}
T^{p}=0={{\omega}^{p}}_{q}{\wedge}{\omega}^{q}+
{{\omega}^{p}}_{z}{\wedge}{\omega}^{z}\label{84}
\end{equation}
\begin{equation}
T^{q}=0={\lambda}e^{-{\frac{\lambda}{2}}z}{\omega}^{z}{\wedge}{\omega}^{q}+{{\omega}^{q}}_{p}{\wedge}{\omega}^{p}+{{\omega}^{q}}_{z}{\wedge}{\omega}^{z}
\label{85}
\end{equation}
and
\begin{equation}
T^{z}=0={{\omega}^{z}}_{p}{\wedge}{\omega}^{p}+{{\omega}^{z}}_{q}{\wedge}{\omega}^{q}
\label{86}
\end{equation}
where $T^{i}$ are the Cartan torsion 2-form which vanishes
identically on a Riemannian manifold. From these expressions one is
able to compute the connection forms which yields
\begin{equation}
{{\omega}^{p}}_{q}=-{\alpha}{\omega}^{p}\label{87}
\end{equation}
\begin{equation}
{{\omega}^{q}}_{z}={\lambda}e^{-{\frac{\lambda}{2}}z}{\omega}^{q}
\label{88}
\end{equation}
and
\begin{equation}
{{\omega}^{z}}_{p}={\beta}{\omega}^{p} \label{89}
\end{equation}
where ${\alpha}$ and ${\beta}$ are constants. Substitution of these
connection form into the second Cartan equation
\begin{equation}
{R^{i}}_{j}={R^{i}}_{jkl}{\omega}^{k}{\wedge}{\omega}^{l}=d{{\omega}^{i}}_{j}+{{\omega}^{i}}_{l}{\wedge}{{\omega}^{l}}_{j}
\label{90}
\end{equation}
where ${R^{i}}_{j}$ is the Riemann curvature 2-form. After some
algebra we obtain the following components of Riemann curvature for
the conformal antidynamo
\begin{equation}
{R^{p}}_{qpq}= {\lambda}e^{-{\frac{\lambda}{2}}z}\label{91}
\end{equation}
\begin{equation}
{R^{q}}_{zqz}= \frac{1}{2}{\lambda}^{2}e^{-{\lambda}z}\label{92}
\end{equation}
and finally
\begin{equation}
{R^{p}}_{zpq}= -{\alpha}{\lambda}e^{-{\frac{\lambda}{2}}z}\label{93}
\end{equation}
We note that only component to which we can say is positive is
${R^{p}}_{zqz}$ which turns the flow stable in this q-z surface.
This component also dissipates away when $z$ increases without
bounds, the same happens with the other curvature components
\cite{21}.
\section{Conclusions}
 In conclusion, we have used a well-known technique to find solutions of Einstein's field equations of gravity
 namely the conformal related spacetime metrics to find a new anti-dynamo solution in MHD three-dimensional Riemannian nonplanar flows.
 Examination of the  Riemann curvature \cite{21} components enable one to
 analyse the stretch and compression of the dynamo flow. New conformal fast dynamo metric
 are obtained from the conformally self-induced equation.It is shown that in the effect of conformal mapping in Riemannian dynamo
 flow is to change the fast dynamo speed. Future perspectives includes the investigation of homological obstructions in the conformal geodesic flows from Anosov flows
  generalizing the investigation of Vishik \cite{22} and Friedlander and Vishik \cite{23}.
\section*{Acknowledgements}
I would like to dedicate this paper to Professor Vladimir I. Arnold
on the ocasion of his senventh birthday. I would like also to thank
CNPq (Brazil) and Universidade do Estado do Rio de Janeiro for
financial supports.

\end{document}